\documentclass[prd
,preprint
,superscriptaddress,showpacs,nofootinbib,%
tightenlines]{revtex4}
\usepackage{epsfig}
\usepackage{color}
\usepackage{amssymb}
\newcommand{\beq}{\begin{equation}}
\newcommand{\eeq}{\end{equation}}
\newcommand{\beqa}{\begin{eqnarray}}
\newcommand{\eeqa}{\end{eqnarray}}
\newcommand{\gA}{\stackrel{\circ}{g}_{\!A}}

\newcommand{\ben}{\begin{displaymath}}
\newcommand{\een}{\end{displaymath}}
\newcommand{\be}{\begin{equation}}
\newcommand{\ee}{\end{equation}}
\newcommand{\bea}{\begin{eqnarray}}
\newcommand{\eea}{\end{eqnarray}}
\begin{document}
\title{Two-nucleon scattering in a modified Weinberg approach with a
  symmetry-preserving regularization}
\author{J.~Behrendt}
 \affiliation{Institut f\"ur Theoretische Physik II, Ruhr-Universit\"at Bochum,  D-44780 Bochum,
 Germany}
\author{E.~Epelbaum}
 \affiliation{Institut f\"ur Theoretische Physik II, Ruhr-Universit\"at Bochum,  D-44780 Bochum,
 Germany}
\author{J.~Gegelia}
\affiliation{Institute for Advanced Simulation, Institut f\"ur Kernphysik
   and J\"ulich Center for Hadron Physics, Forschungszentrum J\"ulich, D-52425 J\"ulich,
Germany}
\affiliation{Tbilisi State  University,  0186 Tbilisi,
 Georgia}
\author{Ulf-G.~Mei\ss ner}
 \affiliation{Helmholtz Institut f\"ur Strahlen- und Kernphysik and Bethe
   Center for Theoretical Physics, Universit\"at Bonn, D-53115 Bonn, Germany}
 \affiliation{Institute for Advanced Simulation, Institut f\"ur Kernphysik
   and J\"ulich Center for Hadron Physics, Forschungszentrum J\"ulich, D-52425 J\"ulich,
Germany}
\affiliation{JARA - Forces and Matter Experiments, Forschungszentrum J\"ulich, D-52425 J\"ulich, Germany}
\affiliation{JARA - High Performance Computing, Forschungszentrum J\"ulich, D-52425 J\"ulich, Germany}
\author{A.~Nogga}
\affiliation{Institute for Advanced Simulation, Institut f\"ur Kernphysik
   and J\"ulich Center for Hadron Physics, Forschungszentrum J\"ulich, D-52425 J\"ulich,
Germany}
\affiliation{JARA - High Performance Computing, Forschungszentrum J\"ulich, D-52425 J\"ulich, Germany}

\begin{abstract}
We consider the nucleon-nucleon scattering problem by applying
time-ordered perturbation theory to the Lorentz invariant formulation
of baryon chiral perturbation theory.  Using a symmetry preserving
higher derivative form of the effective Lagrangian,  
we exploit the freedom of the choice of the renormalization condition
and obtain an integral equation for the scattering amplitude with an
improved ultraviolet behavior.  The resulting formulation is used to
quantify finite regulator artifacts in two-nucleon phase shifts as
well as in the chiral extrapolations of the S-wave scattering lengths
and the deuteron binding energy. This approach can be 
straightforwardly extended to analyze few-nucleon systems and processes
involving external electroweak sources. 
\end{abstract}

\maketitle

\section{\label{introduction}Introduction}

The effective field theory (EFT) approach to the strong
interaction at low energies is a perturbative framework based on an
expansion in terms of small energy and/or masses divided by some large
scale. Higher orders of these small parameters are suppressed at low
energies thus leading to a reliable order-by-order calculation of
physical quantities (see
e.g. Refs.~\cite{Bedaque:2002mn,Bernard:2007zu,Epelbaum:2008ga,Epelbaum:2012vx}
for reviews). An underlying assumption of this approach is that
quantum chromodynamics (QCD) is the correct theory of the strong
interaction. That is, the $S$-matrix of QCD correctly reproduces the
properties of the strong interaction. Based on the symmetries of QCD,
EFT aims at reproducing the $S$-matrix of QCD
in the low-energy region. Poles corresponding to one-particle hadronic states are
represented in an EFT by dynamical fields, branch points are generated by
quantum corrections (i.e. by loop diagrams). Interaction terms of the effective Lagrangian 
generate tree-level diagrams and also contribute in
loop graphs. Some of the interactions leading to contributions
suppressed due to their  higher order  in the  chiral expansion at
tree level can lead to severely divergent contributions in loop
diagrams. However, these large contributions can be removed by
appropriate renormalization.  This program works without complication
in the mesonic sector. Care has to be taken in the single-nucleon sector due
to the non-vanishing chiral limit of the nucleon mass.    
The issue turns out to be highly non-trivial for  few-nucleon systems, 
first considered in the framework of chiral EFT by Weinberg
\cite{Weinberg:rz}.
A complication emerges from the nonperturbative nature of the problem
at hand which requires  infinitely many loop diagrams 
to be  summed up already at leading order (LO). This can be achieved  by
defining an effective potential and solving the Lippmann-Schwinger (LS)
equation for the amplitude. Renormalization of the solutions of such
integral equations is a non-trivial task.  It
led to much controversy and different formulations of the chiral
expansion in the few-nucleon sector 
\cite{Ordonez:1995rz,Birse:1998dk,Gegelia:1998gn,Kaplan:1998tg,Cohen:1998jr,
Fleming:1999ee,Gegelia:2004pz,PavonValderrama:2005gu,Harada:2005tw,   
Nogga:2005hy,Epelbaum:2006pt,Mondejar:2006yu,Soto:2007pg,Beane:2008bt,
Valderrama:2009ei,Yang:2009pn,Soto:2009xy,Epelbaum:2009sd,Birse:2010fj, 
Harada:2010ba,Long:2011xw,
Epelbaum:2012ua,Zeoli:2012bi,Gasparyan:2012km,Harada:2013hwa,
Gasparyan:2013ota,Epelbaum:2013naa,Epelbaum:2014sza}.

In the standard non-relativistic formulation of chiral EFT, the LO LS
equation 
is well known to be linearly divergent due to the singular nature of the
tensor part of the static one-pion exchange potential (OPEP). Accordingly, an iterative
solution of the LS equation in each spin-triplet nucleon-nucleon (NN)
partial wave requires the inclusion of an infinite 
number of counter terms to absorb all emerging ultraviolet (UV)
divergences. Clearly, this is not feasible in practice. The commonly used
approach for dealing with this issue is to introduce a finite UV
cutoff, whose value has to be taken  
of the order of the hard scale of the problem
\cite{Lepage:1999kt,Gegelia:1998iu,Gegelia:2004pz,Epelbaum:2006pt}.   
In practice, the range of such momentum cutoffs turns out to be rather
limited from above due to  
the appearance of spurious deeply bound states, which provide a severe 
(technical) complication for applications beyond the NN system.  

Recently, an alternative approach to chiral EFT for NN scattering has
been formulated \cite{Epelbaum:2012ua}, 
which employs the Lorentz invariant
Lagrangian and makes use of time-ordered perturbation theory
(TOPT). Contrary to the nonrelativistic formulation, it leads to a
renormalizable LO integral equation. Accordingly, the cutoff parameter
can be varied from the hard scale of the problem to infinity. To
achieve renormalizability beyond LO, higher-order corrections to the
amplitude have to be included perturbatively.\footnote{Higher-order
  short-range terms can also be included non-perturbatively, see
  Ref.~\cite{Epelbaum:2015sha} for a calculation of the neutron-proton $^1$S$_0$ phase
shift with the subleading contact term being treated
non-perturbatively.} 
This novel cutoff-independent approach has already been applied to calculate neutron-proton
phase shifts \cite{Epelbaum:2012ua}, perform chiral extrapolations of the S-wave
scattering length and the deuteron binding energy \cite{Epelbaum:2013ij} and to
analyze the electromagnetic form factors of the deuteron
\cite{Epelbaum:2013naa} at LO in the chiral expansion. 
Derivation of the subleading corrections to the NN potential is in progress. 
For recent extensions to $D \bar D^\star$ meson scattering and strangeness
$S=-1$ hyperon-nucleon scattering see Refs.~\cite{Baru:2015tfa} and \cite{Li:2016paq},
respectively. 

As already pointed out, when extending these calculations to higher
orders in the chiral expansion, the corrections to the potential need
to be treated perturbatively if explicit renormalizability of the
scattering amplitude is to be maintained. 
From the practical point of view, it is, however, advantageous to
treat the whole potential including higher-order contributions
non-perturbatively when solving the corresponding integral equations. 
In this case, the general arguments of
Ref.~\cite{Lepage:1999kt,Gegelia:1998iu,Gegelia:2004pz,Epelbaum:2006pt}
apply and the cutoff should not be taken beyond the hard scale of the
problem.  While conceptually equivalent to the standard
non-relativistic chiral EFT with finite cutoff, the Lorentz invariant
approach with non-perturbative treatment of the potential beyond
the LO is expected to offer more flexibility when choosing the cutoff
range.

In this paper we describe in detail the introduction of a
symmetry-preserving higher-derivative regularization using the
chiral EFT formulation for NN scattering of
Ref.~\cite{Epelbaum:2012ua}, see also
Refs.~\cite{Slavnov:1971aw,Djukanovic:2004px,Djukanovic:2006mc} for a related
earlier work along similar lines.\footnote{Our framework differs from the
``semi-relativistic'' scheme of Ref.~\cite{Djukanovic:2006mc} 
by the usage of TOPT \cite{Weinberg:1966jm,Sterman:1994ce}.} 
The main idea of the approach can be explained as follows. 
According to the general formalism of chiral EFT, one has to include
{\it all} terms in the effective Lagrangian which are consistent with
the underlying symmetries of QCD. Some of these terms are redundant in the
sense that one can get rid off them by using field redefinitions.  
Doing so one usually reduces the effective Lagrangian to its minimal
form. 
While convenient, this step is, however, by no means necessary. One
can perform calculations using the original truly most general
effective Lagrangian, or any other ``non-minimal'' form obtained by
applying specifically chosen field redefinitions. Moreover, in certain
cases, it turns out to be more convenient to exploit this freedom of
choosing different forms of the effective Lagrangian. While 
field transformations do not change physical observables when treated
exactly, they correspond to re-summation of certain higher order
contributions in perturbative calculations. In the
context of low-energy EFT, this freedom can be exploited to introduce
a symmetry-preserving regularization
\cite{Slavnov:1971aw,Djukanovic:2004px}, which, unlike  dimensional
regularization, is also applicable beyond standard perturbation
theory.  We use TOPT and define the effective potential as a sum of
all two-particle irreducible diagrams contributing in the scattering
amplitude. The NN scattering amplitude is obtained by
solving the corresponding integral equation. Compared to the work of
Ref.~\cite{Epelbaum:2012ua}, the current approach 
allows one to treat higher-order corrections to the
potential non-perturbatively which has practical advantages when
applying this scheme to three- and more-nucleon systems and 
reactions involving external electroweak probes.  

As an application, we use the resulting formulation to calculate
neutron-proton phase shifts and perform 
chiral extrapolations of NN S-wave scattering lengths and the deuteron
binding energy at LO. The employed higher-derivative regularization
scheme allows us to quantify finite-regulator artifacts in the
calculated observables, see Refs.~\cite{Dyhdalo:2016ygz,Long:2016vnq} for a related
discussion. Our results provide useful information on
the size of the neglected higher-order terms and are confronted with
the findings of Ref.~\cite{Epelbaum:2014efa}, where a different
strategy was used to estimate the theoretical uncertainty. 

Our work is organized as follows: in section~\ref{effective_Lagrangian} we 
consider the effective Lagrangian and
TOPT while the integral equation for the NN scattering amplitude is derived
in section~\ref{intequation}. Next, in section \ref{LOAmplitude}, we
calculate the NN phase shifts at LO and study the dependence of the
chiral extrapolations for S-wave scattering lengths and the deuteron
binding energy on the regulator. Finally, the results of our work are
summarized in  section~\ref{conclusions}.

\section{Formalism}
\label{effective_Lagrangian}

As already pointed out in the introduction, we employ the manifestly
Lorentz invariant effective Lagrangian for pions and nucleons. It
includes the 
purely mesonic, the $\pi {\rm N}$, ${\rm NN}$, $\ldots$ parts,
\begin{equation}
{\cal L}_{\rm eff}={\cal L}_{\pi}+{\cal L}_{\pi {\rm N}}+{\cal
L}_{\rm NN}+\cdots. \label{inlagr}
\end{equation}
The effective Lagrangian is
organised as an expansion in powers of the quark-masses and derivatives. 
The lowest-order mesonic Lagrangian reads
\cite{Gasser:1984yg}\footnote{For the purposes of the current work we
  switch off  the external sources except of course the scalar one that accounts for the explicit symmetry breaking.}
\begin{equation}\label{piLag}
\label{l2} {\cal L}_2=\frac{F^2}{4}\mbox{Tr}\left[\partial_\mu U \left(
\partial^\mu U\right)^\dagger\right] +\frac{F^2 M^2}{4}\mbox{Tr} \left( 
U^\dagger+ U  \right),
\end{equation}
where $U$ is a unimodular unitary $(2\times 2)$ matrix containing
the Goldstone boson fields, $F$ denotes the pion-decay constant in the chiral
limit: $F_\pi=F[1+{\cal O}({m_q})]=92.2$ MeV.
   We consider the isospin-symmetric limit $m_u=m_d={m_q}$.
The lowest-order expression for the squared pion mass is
$M^2=2 B {m_q}$, where $B$ is related to the quark condensate
$\langle \bar{q} q\rangle_0$ in the chiral limit
\cite{Gasser:1984yg}.

The lowest-order Lagrangian of the one-nucleon sector is given as
\cite{Gasser:1988rb}
\begin{equation}
{\cal L}_{\pi {\rm N}}^{(1)}=\bar \Psi \left( i\gamma_\mu D^\mu -m
+\frac{1}{2} \gA\gamma_\mu \gamma_5 u^\mu\right) \Psi,
\label{lolagr}
\end{equation}
where the nucleons are represented by two four-component Dirac
fields $\Psi = (p, \, n )^T$,
$m$ and $\gA$ stand for the nucleon mass and the axial-vector coupling constant in the chiral limit,
respectively, while $D_\mu\Psi = (\partial_\mu +\Gamma_\mu )\Psi $
denotes the covariant derivative. The quantities $u$, $u_\mu$ and $\Gamma_\mu$ are given by
\begin{equation}
u^2=U,\quad u_\mu =iu^{\dagger} \partial_\mu U u^{\dagger},\quad \quad
\Gamma_{\mu}=\frac{1}{2}\left[u^\dagger\partial_{\mu}u
+u\partial_{\mu}u^\dagger \right].
\end{equation}

  To generate an undressed propagator with an improved  ultraviolet
    behavior as compared to that of  
the Lagrangian of Eq.~(\ref{inlagr}), we consider 
an effective Lagrangian with additional terms
\cite{Djukanovic:2004px}. In particular, by 
adding {\it symmetry-preserving} terms to the standard BChPT
Lagrangian of Eq.~(\ref{inlagr}), 
we can obtain the following modified nucleon propagator
\begin{equation}
S_N^\Lambda (p)=\frac{1}{\left(
p\hspace{-.45em}/\hspace{.1em}-m+i0^+\right)} \
\prod\limits_{j=1}^{N_\Psi} \frac{\Lambda_{\Psi j
}^2}{\Lambda_{\Psi j}^2+m^2-p^2-i0^+}. \label{genregfprop}
\end{equation}
Here, the $\Lambda_{\Psi j}$ are, in general, independent parameters.

The choice of the additional terms of the Lagrangian leading to
Eq.~(\ref{genregfprop}) is not 
unique. Following Ref.~\cite{Djukanovic:2004px} for the additional
terms of the Lagrangian we choose 
\begin{equation}
{\cal L}_{\pi N}^{\rm reg}=\sum\limits_{n=1}^{N_\Psi}
\frac{Y_n}{2} \ \left[ \bar \Psi \left( i\gamma_\mu D^\mu -m
\right) \left( D^2+m^2\right)^n \Psi+h.c.\right],\label{Nlagrreg}
\end{equation}
where the $Y_n$ are functions of $\Lambda_{\Psi j}$. For example, for
the modified nucleon propagator
\begin{equation}
S_N^\Lambda (p)=\frac{\Lambda^2}{\left(
p\hspace{-.45em}/\hspace{.1em}-m+i 0^+\right)\left(
\Lambda^2+m^2-p^2-i 0^+\right)} \label{regfprop}
\end{equation}
we have $N_\Psi=1$ and $Y_1=1/\Lambda^2$.

    Depending on the order of the performed calculations
we can choose the modified propagators 
such that all loop Feynman diagrams contributing to the given order
converge.
The additional higher-derivative
terms introduced in the effective Lagrangian can be
re-expressed in a canonical form by using field transformation. This
makes clear that any $\Lambda$ dependence of physical 
quantities can be absorbed in the redefinition of
the parameters of the standard effective Lagrangian. 

    The effective NN Lagrangian contains terms with different numbers of derivatives acting on the nucleon field $\Psi$.
Using field redefinitions and re-organizing the terms one can achieve that the effective Lagrangian contributing to the
LO NN potential contains only terms without derivatives:
\begin{eqnarray}
{\cal L}_{\rm NN} &=& C_S^a \ \bar\Psi
\tau^a\Psi \ \bar\Psi\tau^a \Psi + C_{T}^a \
\bar\Psi\tau^a\sigma_{\mu\nu} \Psi \
\bar\Psi\tau^a\sigma^{\mu\nu}\Psi
+ C_{AV}^a \bar\Psi\tau^a\gamma_5 \gamma_\mu\Psi \
\bar\Psi\tau^a\gamma_5\gamma^\mu\Psi  \nonumber \\
&+& C_V^a \bar\Psi\tau^a
\gamma_\mu \Psi \ \bar\Psi\tau^a \gamma^\mu \Psi, \label{NNLagrdreg}
\end{eqnarray}
where summation over $a=0,1,2,3$ is implied, with $\tau^0$  denoting
the unit matrix and $\tau^i$
($i=1,2,3$) referring to  the  Pauli (isospin) matrices. Further, $C_I^1=C_I^2=C_I^3$
for all $I$. Note that the LO effective Lagrangian of
Eq.~(\ref{NNLagrdreg}) also contributes to the potential at higher
orders. 
For the most general Lorentz invariant NN Lagrangian of the second
order  (in small momentum and quark masses) see
Ref.~\cite{Girlanda:2010ya}. 

\medskip


For the purposes of the current work it is convenient to take the
additional terms of the following form 
\begin{eqnarray}
\label{NlagrregTW}
{\cal L}_{\pi N}^{\rm reg} &=& 
\frac{1}{2\Lambda^2} \ \left[ \bar \Psi \left( i\gamma_\mu D^\mu -m
\right) \left( D^2+m^2\right) \Psi+h.c.\right] \\
&\equiv& -\frac{1}{2\Lambda^2} \ \left[ \bar \Psi \left( i\gamma_\mu D^\mu -m
\right) \vec D^2 \Psi+h.c.\right]
+ \frac{1}{2\Lambda^2} \ \left[ \bar \Psi \left( i\gamma_\mu D^\mu -m
\right) \left( D_0^2+m^2\right) \Psi+h.c.\right] .
\nonumber
\end{eqnarray}
Here and in the following, we include the contribution of the first
term of Eq.~(\ref{NlagrregTW}) to the nucleon two-point function
non-perturbatively, which affects the form of the propagator, while  
the second term is treated in perturbation theory.  
The corresponding propagator has the form 
\begin{equation}
S_N^\Lambda (p)=\frac{\Lambda^2}{\left(
p\hspace{-.45em}/\hspace{.1em}-m+i \epsilon \right)\left(
\Lambda^2+\vec p\, ^2\right)}. \label{regfpropTW}
\end{equation}

To obtain the rules of the TOPT  we first draw all Feynmann diagrams,
relevant for the given process (in principle an infinite number of
them), assign the momenta to propagator lines and perform the trivial
integrations using the delta functions appearing at the  vertices. The
remaining overall delta function ensures the momentum
conservation for the external legs of diagrams. Next we perform 
integrations over the zeroth components of the loop integration
variables. As a result we are led to the diagrams of the TOPT \cite{Weinberg:1966jm,Sterman:1994ce}. 
The same time-ordered diagrams can be obtained directly using the
following rules of the TOPT:\footnote{As our EFT Lagrangian contains
  fermion fields and the interaction terms with time derivatives, we
  need to keep track of the poles for which the residues have been
  picked up.} 

\medskip

\noindent
The $S$ matrix for a transition $\alpha\to\beta$ may be written as
\begin{equation}
S_{\beta\alpha}=\delta_{\beta\alpha}-(2 \pi)^4
i\,M_{\beta\alpha}\delta^4(P_\beta-P_\alpha)\Pi_n^{\alpha,\beta}(2
\pi)^{-3/2}(2 \omega_n)^{-1/2}, 
\label{Smatrix}
\end{equation}
where $P^\mu$ is the total four-momentum, $\omega_n=\left(\vec p_n{
  }^2+m_n^2\right)^{1/2}$, where $\vec p_n$ is the three-momentum of a
particle  with mass $m_n$ and $\Pi_n^{\alpha,\beta}$ stands for the product over all particles in
the initial and final states. The  invariant amplitude $M$ is obtained by
using the following diagrammatic rules: 

\begin{itemize}
\item
Draw all possible time-ordered diagrams for the transition
$\alpha\to\beta$. That is, draw each Feynmann diagram with $N$ vertices
$N!$ times while ordering the vertices in every possible way in a sequence
running from right to left.  Label each line with a four dimensional
momentum $p=(p_0,\vec p)$ as prescribed by the corresponding Feynmann
diagram. 
\item
Include a factor
\begin{equation}
(2 \pi)^{-3}(2 \omega)^{-1}
\label{efactor}
\end{equation}
for every internal line. For fermion lines, include also factors
$(p\hspace{-.4em}/\hspace{.1em}+m)/(\vec p\, ^2+\Lambda^2)$. 
\item
Multiply the expressions with the coupling constants and include the
relevant factors such as momenta, $\gamma$-matrices etc.~for every vertex. 
For  each time-ordered diagram, 
the zeroth component of the integration variable in the numerators of
fermion propagators is to be replaced by
the energy of the corresponding fermion line. The sign of this energy
is positive for a particle line  and negative for an antiparticle line.
Care has to be taken also of zeroth components present in interaction vertices, 
 that is, one needs to identify the poles in the complex plane of the zeroth components of the
  integration momenta, which have been picked up to obtain 
the given TOPT diagram, and substitute the corresponding expressions or the zeroth components of
the momenta in the vertices.
\item
For every intermediate state $\gamma$, i.e.~a set of lines between any
of two vertices, include an energy denominator 
\begin{equation}
\left[E_\alpha-E_\gamma+i\,\epsilon\right]^{-1},
\label{denominator}
\end{equation}
where $E=\sum \omega$ is the total energy of the state.
\item
Integrate the product of these factors over all internal (three)
momenta, and sum the result over all diagrams. 
\end{itemize}

\section{Integral equation for the off-shell scattering amplitude}
\label{intequation}

The N-nucleon scattering amplitude $M_{2N}$ is obtained from 2N-point
vertex function $\tilde\Gamma_{2N}$ by applying the standard LSZ
formula 
\begin{equation}
M_{2N}=Z_\psi^N\, \bar u_{out1}\ldots \bar u_{outN}\,\tilde\Gamma_{2N}\,u_{in 1}\ldots u_{inN}, \label{LSZtosh}
\end{equation}
where $Z_\psi$ is the residue of the nucleon propagator and $u_{in
  I}$, $\bar u_{out J}$ are Dirac spinors of the incoming and outgoing
nucleons. 
The vertex function with on-shell momenta is given as a sum of an infinite number of TOPT diagrams.

 For nucleon-nucleon scattering, the purely two-nucleon intermediate states are enhanced \cite{Weinberg:rz}. Therefore it is convenient to
define  effective potentials as the sums of all two-particle irreducible TOPT diagrams.
The on-shell four-point vertex function $\tilde \Gamma$ is then given by an infinite series
\begin{eqnarray}
\tilde \Gamma &=& \tilde V+\bar V G\, \bar V +\bar V G\, V G\, \bar V+ \bar V G\, V G\, V G\, \bar V +\cdots \nonumber\\
& = & \tilde V+\bar V G\, \bar V +\bar V G\, \left[ V + V G\, V  +\cdots \ \right] G\, \bar V = \tilde V+\bar V G\, \bar V +\bar V G\,\Gamma G\, \bar V,
\label{Tseries}
\end{eqnarray}
where $G$ is the two-nucleon Greens function and $\tilde \Gamma$, $\Gamma$, $\tilde V$, $\bar V$ and $V$  are the on-shell vertex function, the off-shell vertex function, the on-shell potential, the half-off shell potential and the off-shell potential,   respectively.
Note here that we include the numerators of the fermion propagators in the definition of the potential so that $G$ contains a factor of $2 m$ for each nucleon.
The off-shell vertex function $\Gamma$ can be obtained by solving the following equation:
\begin{equation}
\Gamma=V+ V G\,\Gamma\,.
\label{Teq1}
\end{equation}
We define a projected potential
\begin{equation}
V_P=P_+ P_+\, V \,P_+ P_+,
\label{Vprojected}
\end{equation}
where
\begin{equation}
P_+ = \frac{1+v \hspace{-.45em}/\hspace{.1em}}{2}, \ \ \ 
v=(1,0,0,0),
\label{spellout}
\end{equation}
and assume that it can be expanded in some generic parameter according to 
\begin{equation}
V_P=  V_{P0}+ V_{P1}+\cdots \,.
\label{Vexp}
\end{equation}
Next, we introduce the corresponding vertex function $\Gamma_0$ as the solution to the following equation
\begin{equation}
\Gamma_0= V_{P0}+V_{P0}\, G\,\Gamma_0 \,.
\label{LOgeqPTilde}
\end{equation}
It is easily seen from Eq.~(\ref{LOgeqPTilde}) that
\begin{equation}
\Gamma_0=\Gamma_{P0}\equiv P_+ \ldots P_+\, \Gamma_0 \,P_+ \ldots P_+.
\label{Gamma0equiv}
\end{equation}
Within our approach we identify the reduced LO potential $V_{P0}$ as the LO effective potential $V_0$ and expand the effective potential as
\begin{equation}
V \equiv V_0+(V-V_0) = V_0 +V_1+V_2 +\cdots.
\label{Veffexp}
\end{equation}
We treat $V_0$ non-perturbatively, i.e. we calculate the LO vertex function $\Gamma_0$ from Eq.~(\ref{LOgeqPTilde}) and include the corrections perturbatively.
The NLO vertex function $\Gamma_1$ is given as
\begin{equation}
\Gamma_1 = V_1+V_1 G \,\Gamma_0+\Gamma_0 G \,V_1 +\Gamma_0 G \,V_1 G \,\Gamma_0.
\label{NLOVF}
\end{equation}
It is straightforward to express the further higher-order corrections
to the vertex function in terms of the
potentials of the corresponding orders and vertex functions of lower orders.
Further, when calculating the effective potential, we write the numerator of
the standard fermion propagator  as 
\begin{equation}
p\hspace{-.45em}/\hspace{.1em}+m 
= 2 m\, P_+   + \left(p
\hspace{-.4em}/\hspace{.1em}-m\,v \hspace{-.45em}/\hspace{.1em}\right),
\label{Sfpexpanded}
\end{equation}
and identify the second term as a higher-order correction.

To determine the physical amplitude order by order in an expansion in
terms of small parameters we rewrite the Dirac spinors as 
\begin{eqnarray}
u & = & \left(1+\frac{p
\hspace{-.4em}/\hspace{.1em}-p\cdot v}{m+p\cdot v}\right) P_+\, u(p) = u_0 + u_1.\nonumber
\\
\bar u & = & \bar u(p) P_+ \left(1+\frac{p
\hspace{-.4em}/\hspace{.1em}-p\cdot v}{m+p\cdot v}\right) =\bar u_0 +\bar u_1\,.
\label{deqsolexpanded}
\end{eqnarray}
For small three-momenta of the nucleons, we identify $u_1$ and $\bar u_1$
as corrections suppressed compared to the dominant $u_0$ and $\bar
u_0$ terms. 

For two-nucleon scattering, we calculate the scattering amplitude in
the center-of-mass frame. 
The vertex function $\Gamma\left(E,
\vec p\,',\vec p\right)$ satisfies the equation (the spin and isospin
indices are suppressed) 
\begin{eqnarray}
\Gamma\left(E,
\vec p\,',\vec p\right)&=& V\left(E,
\vec p\,',\vec p\right) - \frac{m^2}{2}\int \frac{d^3 \vec k}{(2
                           \,\pi)^3} \  V\left(E, 
\vec p\,',\vec k\right)  
G(E,k,\Lambda)
   \,\Gamma\left(E,
\vec k,\vec p\right), \nonumber
\\
G(E,k,\Lambda) &=&  
\frac{\Lambda^4}{(\vec k^2+\Lambda^2)^2 (\vec k^2+m^2)\Big(
E/2-\sqrt{\vec k^2+m^2}+i \epsilon \Big)} .
\label{MeqLOk0integratedHDR}
\end{eqnarray}
Here,  $\vec p$ and  $\vec p\,'$ are the three-momenta of the incoming and
outgoing nucleons, respectively, and $E=2 \sqrt{\vec q\,^2+m^2}$ is
the energy of  the two nucleons.  
\begin{figure}
\vspace{-2.cm}
\includegraphics[width=0.78\textwidth,keepaspectratio,angle=0,clip]{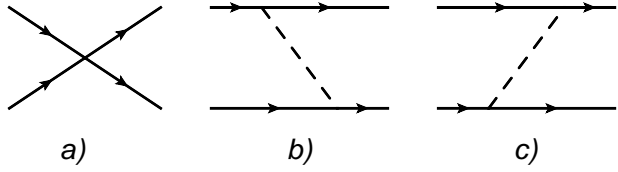}
\vspace{-12.cm}
\caption[]{\label{NNfig4:fig} Time-ordered diagrams contributing to
  the LO NN potential.}
\end{figure}
In the partial wave basis, Eq.~(\ref{MeqLOk0integratedHDR}) leads to
the following  equations with the partial wave projected potential
$V^{sj}_{l'l}\left(p',p\right)$, 
\begin{eqnarray}
T^{sj}_{l'l}\left(E, {p'}
,p\right) = V^{sj}_{l'l}\left(E,p',p\right) + \frac{m^2}{2} \sum_{l''}\int_0^\infty
\frac{d k\,k^2}{(2 \pi)^3} V^{sj}_{l'l''}\left(E,p',k\right) G(E,k,\Lambda) T^{sj}_{l''l}
\left(E, k ,p\right).
\label{PWEHDR}
\end{eqnarray}
A standard UV counting shows that in the limit of large
integration momenta, Eq.~(\ref{MeqLOk0integratedHDR}) 
has a milder UV behavior than the corresponding LS equation.
That is, its solutions are expected to show less sensitivity to the variation of the
finite values of the parameter $\Lambda$.

\section{Applications to Nucleon-nucleon scattering at Leading Order}
\label{LOAmplitude}

The LO effective potential for NN scattering, 
$ V_0 =V_{0,C} + V_{0,\pi}$,
consists of the contact interaction part and two one-pion exchange  time-ordered
diagrams shown in Fig.~\ref{NNfig4:fig}.  The
projected OPEP has the  form
\begin{eqnarray}
V _{\pi}  \left(
\vec p\,',\vec p\right) & = & -\frac{g_A^2}{4\,F^2}\,\frac{\
\vec \tau_1\cdot\vec\tau_2}{\sqrt{\left(
\vec p-\vec p\,'\right)^2+M^2}}\nonumber\\
&\times& \frac{\left[
\vec\sigma_1\cdot \left(\vec p-\vec p\,'\right)\right]\left[
\vec\sigma_2\cdot \left(\vec p-\vec p\,'\right)\right]}{\sqrt{\left(
\vec p-\vec p\,'\right)^2+M^2}+\sqrt{
\vec p\,^2+m^2}+\sqrt{\vec p\,'^2+m^2}-E-i\,\epsilon }\,.
\label{opepotreduced}
\end{eqnarray}
In the calculations of the phase shifts described below, we
approximate the two-nucleon energy $E$  in Eq.~(\ref{opepotreduced})
by $2 m$.  

We solve the integral equation
(\ref{PWEHDR}) using standard numerical methods. For the various
parameters, we employ the values of 
$M_\pi = 138.0$~MeV, $F_\pi = 92.2$~MeV and $m=938.9$~MeV.  Further,
for the nucleon axial-vector coupling, we
use the effective value $g_A =1.285$ which takes into account the
Goldberger-Treiman discrepancy. Two linear combinations of low-energy constants $C_S+C_V$ and
$C_{AV}+2C_T$ contributing at LO \cite{Djukanovic:2006mc} are adjusted to reproduce the $^1$S$_0$ and $^3$S$_1$ 
phase shifts from the Nijmegen
partial wave analysis \cite{Stoks:1993tb} below $E_{\rm lab} = 50$~MeV.   

\begin{figure}
\includegraphics[width=\textwidth,keepaspectratio,angle=0,clip]{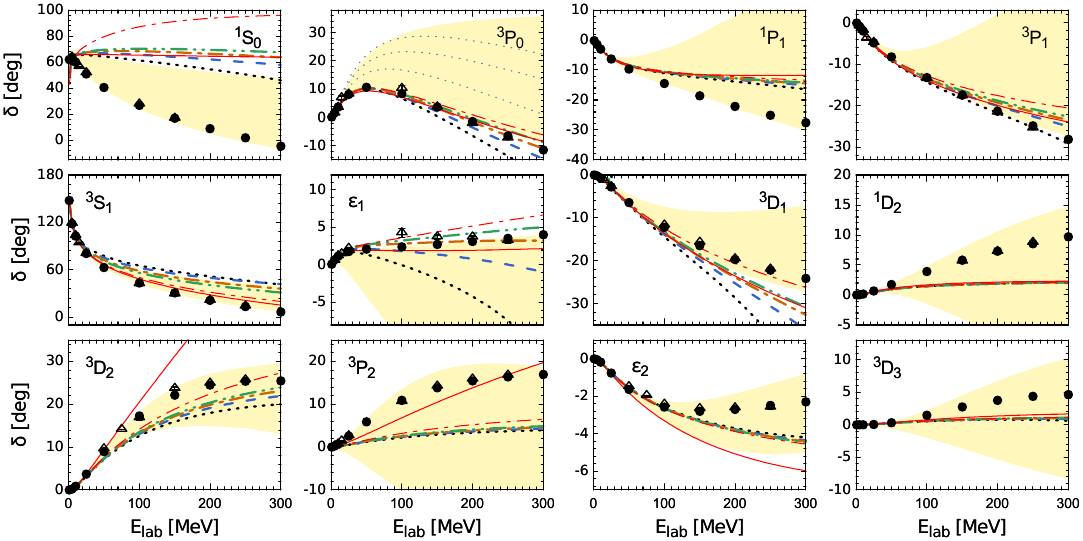}
\caption[]{\label{NNfig4x:fig} 
(Color online) Phase shifts and mixing angles calculated at LO as functions of laboratory energy in
comparison with the Nijmegen \cite{Stoks:1993tb} (filled circles) and
GWU single-energy partial wave analyses \cite{VT} (open triangles).  
Thick black dotted, blue dashed, brown dashed-dotted and green dashed-double-dotted lines  
correspond to $\Lambda=600$,  $800$, $1000$ and $1200$~MeV,
respectively. In the $^3$P$_0$ channel, the three upper curves shown
by thin violet dotted
lines show the predictions
including the infinitely strong repulsive contact
interaction for $\Lambda=800$, $1000$ and $1200$~MeV (from bottom to top). 
Thin red dashed-dotted lines give the result for $\Lambda \to
\infty$, while thin solid lines show the LO results from
Ref.~\cite{Epelbaum:2012ua} obtained in the limit of  $\Lambda \to
\infty$ by employing  the \emph{static} form of the one-pion exchange
potential. Light shaded areas depict the estimated theoretical
uncertainty from the truncation of the chiral expansion at LO from
Ref.~\cite{Epelbaum:2014efa} for the regulator choice of
$R=0.9$~fm. 
}
\end{figure}

In Fig.~\ref{NNfig4x:fig}, 
we show our results for the neutron-proton S-, P- and
D-waves  and the mixing angles $\epsilon_1$ and $\epsilon_2$ for the
regulator values of the order of the expected breakdown scale in the
problem, namely $\Lambda = 600$, $800$, $1000$ and $1200$~MeV,   
in
comparison with the Nijmegen \cite{Stoks:1993tb} and  the GWU single-energy
partial wave analysis (PWA) \cite{VT}. We observe a strong $\Lambda$-dependence
in the $^3P_0$ partial wave caused by the presence of a
pole in the complex plane on the unphysical sheet. 
To get the phase shifts closer to
the PWA we have included the NLO contact interaction.\footnote{We
remind the reader that the contributions of a given term in the
effective Lagrangian to observables depend on the employed scheme and
the choice of renormalization conditions. The results of Refs.~\cite{Epelbaum:2014efa,Epelbaum:2014sza} based
on the nonrelativistic framework and utilizing a coordinate-space
regulator chosen in the range of $R=0.8 \ldots 1.2$~fm do not support
the need to depart from Weinberg's original power counting for contact
interactions, see also \cite{Epelbaum:2015pfa} for a related discussion.} 
As the OPEP is strongly attractive in this
channel, the introduced counter term has to be repulsive. Notice
that for the employed form of the integral equation and
higher-derivative regularization, we would have to take an infinitely
large repulsive value for the counter
term already for $\Lambda\sim 566$ MeV to reproduce the low-energy
behavior of the phase shift. For larger values of $\Lambda$, the low energy phase shifts can
be reproduced by adjusting an {\it attractive} counter term, which,
however, leads to the appearance of  a deeply bound state in this
channel. We also show in Fig.~\ref{NNfig4x:fig} the $^3$P$_0$ phase
shifts which result from taking an infinitely large repulsive value
for the counter term instead of tuning it to the Nijmegen PWA, which
prevents the appearance of deeply-bound states in the considered
range of $\Lambda$. 

For the sake of comparison, we also
show in Fig.~\ref{NNfig4x:fig} the results from Ref.~\cite{Epelbaum:2012ua} which
correspond to the limit of $\Lambda \to \infty$. As already pointed
out in the introduction, such a limit can be taken in the LO equation,
but does not represent a legitimate procedure for a non-perturbative
inclusion of higher-order corrections.  Notice that our
finite-$\Lambda$ predictions are, strictly speaking, not expected to converge
against the infinite-$\Lambda$ results of
Ref.~\cite{Epelbaum:2012ua} shown by the thin solid lines due to the different form of the employed
one-pion exchange potential. 
In particular, a static
approximation of the nonstatic expression for the OPEP in Eq.~(\ref{opepotreduced}) was employed in that
work. Clearly, both choices for the OPEP are valid as the difference
between the two forms is of a  higher order. Our
infinite-$\Lambda$ predictions based on the non-static form of the
OPEP given in Eq.~(\ref{opepotreduced}) are shown by the thin
dashed-dotted lines in Fig.~\ref{NNfig4x:fig}.  

We are now in the position to discuss the implications of our findings
in connection with the estimation of the theoretical uncertainty at
LO. To this aim, we show by the light-shaded areas the LO predictions
of the new generation of chiral NN potentials of
Ref.~\cite{Epelbaum:2014efa} 
obtained using a semi-local regularization within the nonrelativistic
framework, 
along with the estimated uncertainty from
the truncation of the chiral expansion, see that reference for more
details.  We restrict ourselves to showing the results corresponding
to the regulator choice of $R=0.9$~fm found to lead to the most accurate
predictions. Notice that the theoretical uncertainty was quantified in
that work by estimating the size of higher-order contributions to the
potential \emph{without} relying on the variation of the regulator
$R$. Thus, the results of our study utilizing a large
variation of the cutoff $\Lambda$ in the range from $600$~MeV  to
$\infty$ provide an excellent opportunity to test the
reliability of the uncertainty quantification approach formulated in
Ref.~\cite{Epelbaum:2014efa}.  

It is reassuring to see that our
predictions for different values of $\Lambda$ shown by various lines
in Fig.~\ref{NNfig4x:fig} lie within the
estimated theoretical uncertainty in almost all considered cases. The
most notable exception is
the $^1$S$_0$ phase shift, for which we do observe a somewhat larger
deviation from the Nijmegen PWA for $\Lambda = \infty$ as compared with the results of
Ref.~\cite{Epelbaum:2014efa}.  Notice, however, that the deviations between our
results for $\Lambda = 600 \ldots 1200$~MeV from the Nijmegen phase
shifts are comparable to the width of the uncertainty band. 
It is well known that the one-pion exchange
potential only generates about a half of the effective range in this
channel, see e.g. Ref.~\cite{Baru:2015ira}, and that the contribution of the subleading
contact interaction is large \cite{Epelbaum:2015sha}. These features
are, in fact, 
reflected in the rather large estimated theoretical uncertainty in
this channel. The dominant role of short-range interactions in this
partial wave is also consistent with by far the slowest observed convergence of
the calculated phase shifts 
with respect to
$\Lambda$.
Another quantity which appears to be rather sensitive to short-range
physics is the mixing angle $\epsilon_1$. This observation  is consistent with
results obtained within different approaches, see
e.g. Refs.~\cite{Epelbaum:2015sha,Gezerlis:2014zia,Marji:2013uia}.  

Our results also illustrate, that the assessment of the
theoretical uncertainty based on a cutoff variation alone should be
taken with care as it is insensitive to neglected long-range
interactions. In particular, for the nonstatic OPEP in Eq.~(\ref{opepotreduced}), we
obtain nearly cutoff-independent results for
the $^1$P$_1$, $^1$D$_2$, $^3$P$_2$ and $^3$D$_3$ phase
shifts and the mixing angle $\epsilon_2$ over the range of $ 600 \mbox{ MeV} <
\Lambda < \infty$, which, however, deviate from the empirical phase
shifts. This lets one conclude that the discrepancies with the Nijmegen PWA
in these channels are largely driven by higher-order long-range
interactions. Employing the two different forms of the OPEP which are
equivalent at LO allows us to probe sensitivity to certain kinds of
higher-order  long-range contributions and leads to a more realistic
uncertainty estimation in the  $^3$P$_2$ partial wave and 
the mixing angle $\epsilon_2$. Generally, the results of the present
work fit well with those of the nonrelativistic calculation in
Ref.~\cite{Epelbaum:2015sha} and show no contradictions with the theoretical
uncertainty at LO estimated in that paper. 

\begin{figure}
\includegraphics[width=\textwidth,keepaspectratio,angle=0,clip]{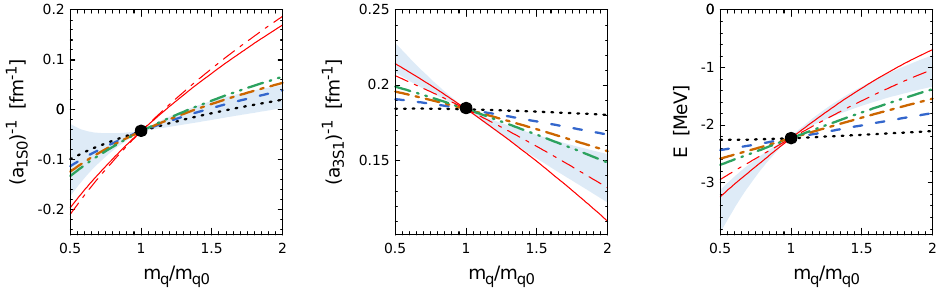}
\caption[]{\label{aInvBe} (Color online) Quark mass dependence of the inverse 
  scattering lengths $a_{^1S_0}^{-1}$,
$a_{^3S_1}^{-1}$ and the deuteron binding energy $E$ at LO. Solid dots show
the empirical values for the considered observables at the physical
value $m_{q0}$ of the quark mass while light-shaded bands give the N$^2$LO
results from Ref.~\cite{Berengut:2013nh} based on the resonance saturation hypothesis for
short-range operators. For remaining notation see Fig.~\ref{NNfig4x:fig}. 
}
\end{figure}

As a second application, we consider the quark- or, equivalently, pion-mass dependence of the
neutron-proton S-wave scattering lengths $a_{^1S_0}$, $a_{^3S_1}$ and the deuteron binding
energy. At LO in the modified Weinberg approach, the $M_\pi$-dependence of the
scattering amplitude emerges entirely from the explicit
$M_\pi$-dependence of the pion propagator in the OPEP and, therefore, 
can be predicted in a parameter-free way. In Ref.~\cite{Epelbaum:2013ij}, we have
shown the predicted $M_\pi$-dependence of $a_{^1S_0}^{-1}$,
$a_{^3S_1}^{-1}$ and the deuteron binding energy $E$ for $\Lambda \to \infty$
using the static form of the OPEP. For the spin-triplet channel, the
predicted $M_\pi$-dependence suggests a weaker attraction for 
larger-than-physical values  of the quark masses in agreement with the
results of Refs.~\cite{Flambaum:2007mj,Berengut:2013nh}. On the other hand, lattice QCD
calculations for pion masses $M_\pi > 300$~MeV indicate an opposite
trend with the deuteron being stronger bound
\cite{Beane:2011iw,Yamazaki:2015asa,Orginos:2015aya,Beane:2013br,Yamazaki:2012hi},
except for the results by HAL QCD collaboration
which finds no bound deuteron for 
$M_\pi = 469 \ldots 1171$~MeV~\cite{Inoue:2011ai}. For a summary of the current
status of lattice QCD results and a related discussion based on low-energy theorems for NN
scattering see Refs.~\cite{Baru:2015ira,Baru:2016evv}. 
Thus, it is
important to quantify the theoretical uncertainty of the LO prediction
of Ref.~\cite{Epelbaum:2013ij}. To this aim, we calculate the quark
mass dependence of  $a_{^1S_0}^{-1}$,
$a_{^3S_1}^{-1}$ and $E$ at LO using the higher-derivative regularization
framework for different values of $\Lambda$ as visualized in
Fig.~\ref{aInvBe}. 
Here and in what follows, we work in the isospin
limit and employ the LO approximation to relate
the pion and the average  light-quark masses, $M_\pi$ and $m_q$, namely,
$M_\pi^2 = 2 B m_q$.  Furthermore, in all cases discussed below, we
tune the contact interaction to exactly reproduce the corresponding
observable at the physical point. 

For all considered quantities, the difference in the predictions based
on the static and nonstatic forms of the OPEP in the
infinite-$\Lambda$ limit turns out to be small showing that the corresponding
higher-order long-range terms in the potential play a minor role. On
the other hand, we do observe a significant
$\Lambda$-dependence indicating that the contributions of the
neglected quark-mass dependent contact interactions 
are substantial, see also \cite{Berengut:2013nh,Epelbaum:2002gb,Epelbaum:2002gk,Beane:2002vs,Beane:2002xf}
for similar conclusions achieved using different approaches. 
For the spin-singlet channel, the resulting $\Lambda$-dependence of the
chiral extrapolation of the scattering length for unphysical quark masses shows a similar pattern
to the scattering phase shift at the physical value of the quark
mass. In particular, we observe a very slow convergence with respect to
the cutoff $\Lambda$. Interestingly, for $a_{^1S_0}$, our results for $\Lambda$ in the
range of $\Lambda = 600 \ldots 1200$~MeV appear to be in a very good agreement
with the predictions of Ref.~\cite{Berengut:2013nh} obtained within the standard
nonrelativistic formulation at N$^2$LO in the chiral expansion. In
that work, the $M_\pi$-dependence of the short-range operators was
estimated via the resonance saturation hypothesis 
by invoking unitarized chiral perturbation theory in combination with
lattice QCD results to describe the dependence of the resonance
positions on the quark mass. The light-shaded bands shown in
Fig.~\ref{aInvBe} correspond to cutoff variation over a
certain range, see \cite{Berengut:2013nh} for more details, but do not
take into account the uncertainty associated with the usage of the
resonance saturation hypothesis. Interpreting the spread between the
various lines in Fig.~\ref{aInvBe} as an estimation of  the
contributions from higher order $M_\pi$-dependent 
contact interactions, we conclude that the LO predictions cannot
exclude the possibility of an 
increasing attraction in the $^3$S$_1$ channel for unphysically
large values of $M_\pi$ reported in the recent lattice QCD
calculations. Notice, however, that the preferred scenario based on
our LO calculations still
corresponds to decreasing of the deuteron binding energy with
increasing values of the pion mass near the physical point.

\section{Summary}
\label{conclusions}

We considered the nucleon-nucleon scattering problem in the framework
of higher derivative formulation of BChPT of
Ref.~\cite{Djukanovic:2004px} by applying the TOPT to the Lorentz
invariant effective Lagrangian. 
The formulation preserves all
underlying symmetries and is therefore
well suited for studying processes involving external electroweak probes and
chiral extrapolations. 
At the same time, it leads to 
non-singular integral equations for the NN scattering amplitude at
any given order in the chiral expansion and, differently to the
framework of Ref.~\cite{Epelbaum:2012ua}, allows for a  non-perturbative inclusion of
higher-order contributions to the potential. This feature may be
particularly useful for facilitating the generalization of this approach to
three- and more-nucleon systems and is achieved by making use of the freedom
in the choice of the form of the 
effective Lagrangian or, equivalently, renormalization conditions
parametrized in terms of the scale $\Lambda$.   
Given that the LO equation for the NN scattering amplitude is
renormalizable in our scheme, the parameter $\Lambda$ can be varied over a large
range or even completely eliminated yielding the approach of
Ref.~\cite{Epelbaum:2012ua}. On the other hand, if higher-order contributions to the
potential are to be taken into account non-perturbatively, the
parameter $\Lambda$ needs to  be chosen of the order of the expected 
breakdown scale of the theory such as e.g. the mass of the $\rho$
meson.  

We have applied this framework to test the estimation of the
theoretical uncertainty  for NN phase shifts and mixing
angles of Ref.~\cite{Epelbaum:2014efa} at LO and to quantify the accuracy of the
chiral extrapolations of the S-wave scattering lengths $a_{^1S_0}$, $a_{^3S_1}$ and the
deuteron binding energy $E$ predicted in Ref.~\cite{Epelbaum:2013ij}. The
resulting sensitivity of the NN scattering observables to the considered
variation of the parameter $\Lambda$ over a large range is found to be
consistent with the LO uncertainty bands of Ref.~\cite{Epelbaum:2014efa}
generated without relying on cutoff variation. For the
chiral extrapolations of $a_{^1S_0}$, $a_{^3S_1}$ and $E$, we find a rather sizable sensitivity to 
neglected higher-order $M_\pi$-dependent contact interactions which, however, 
should not come as a surprise given the strongly fine-tuned nature of these quantities, 
also noted in earlier work
\cite{Berengut:2013nh,Epelbaum:2002gb,Epelbaum:2002gk,Beane:2002vs,Beane:2002xf}.

\acknowledgments

This work was supported in part by Georgian Shota Rustaveli National
Science Foundation (grant FR/417/6-100/14) and by DFG (SFB/TR 110,
``Symmetries and the Emergence of Structure in QCD'') and BMBF (contract No. 05P2015 -
NUSTAR R\&D). 
The work of UGM was supported in part by The Chinese Academy of Sciences 
(CAS) President's International Fellowship Initiative (PIFI) grant no. 2015VMA076.
Parts of the numerical calculations have been performed on JURECA and JUQUEEN
of the J\"ulich supercomputing centre, J\"ulich, Germany.

\end{document}